\documentclass[pdflatex,sn-basic, iicol]{sn-jnl}

\usepackage{graphicx}%
\usepackage{multirow}%
\usepackage{amsmath,amssymb,amsfonts}%
\usepackage{amsthm}%
\usepackage{mathrsfs}%
\usepackage[title]{appendix}%
\usepackage{xcolor}%
\usepackage{textcomp}%
\usepackage{manyfoot}%
\usepackage{booktabs}%
\usepackage{algorithm}%
\usepackage{algorithmicx}%
\usepackage{algpseudocode}%
\usepackage{listings}%

\usepackage{breakurl}
\usepackage{enumitem}

\theoremstyle{thmstyleone}%
\theoremstyle{thmstyletwo}%

\theoremstyle{thmstylethree}%

\begin{document}

\title[Dialogue Agents that Share Family Information to Strengthen Grandparent--Grandchild Relationships]{Dialogue Agents that Share Family Information to Strengthen Grandparent--Grandchild Relationships}


\author*[1]{\fnm{Seiya} \sur{Mitsuno}}\email{mitsuno.seiya.es@osaka-u.ac.jp}

\author[2]{\fnm{Midori} \sur{Ban}}\email{ban-m@tachibana-u.ac.jp}

\author[1]{\fnm{Hiroshi} \sur{Ishiguro}}\email{ishiguro@irl.sys.es.osaka-u.ac.jp}

\author[1]{\fnm{Yuichiro} \sur{Yoshikawa}}\email{y.yoshikawa.es@osaka-u.ac.jp}

\affil*[1]{\orgdiv{Graduate School of Engineering Science}, \orgname{The University of Osaka}, \orgaddress{\street{1-3, Machikaneyama-cho}, \city{Toyonaka}, \postcode{560-8531}, \state{Osaka}, \country{Japan}}}

\affil[2]{\orgdiv{Faculty of Engineering}, \orgname{Kyoto Tachibana University}, \orgaddress{\street{34 Oyakiyamada-cho}, \city{Kyoto}, \postcode{607-8175}, \state{Kyoto}, \country{Japan}}}


\abstract{
Social isolation among older adults has become a critical concern, as reduced opportunities for conversation and weakened family relationships negatively affect mental health. This study proposes a dialogue agent that supports older adults by fostering both a relationship with the agent and a relationship with their grandchild through sharing everyday information. The agent operates on a chatbot platform and engages in daily conversations with older adults and their grandchildren, exchanging information gathered from each party to enhance conversational engagement and social connection. We conducted a ten-day empirical experiment with 52 grandparent--grandchild pairs. The results suggest that older adults became more willing to interact with the proposed agent, which shared information about their grandchildren, and that the psychological connection between grandparents and grandchildren was strengthened. Furthermore, daily interactions with the agent were associated with reduced anxiety in both older adults and their grandchildren. These findings indicate that a dialogue agent that shares personal information can be an effective approach to supporting older adults by simultaneously offering conversational opportunities and promoting family connectedness. Overall, this study provides valuable insights into the design of dialogue agents that effectively address social isolation among older adults.}




\maketitle

\section{Introduction} \label{sec:intro}
In recent years, social isolation among older adults has emerged as a serious concern due to its negative effects on mental health \citep{Holt-Lunstad2024-od}. Social isolation is defined as a lack of social interaction and connection with others \citep{Valtorta2012-lo, Nakou2025-jz}, and is often caused by reduced opportunities for daily conversation and weakened ties with family, neighbors, and local communities \citep{Donovan2020-tc, Suragarn2021-uz}. This issue is particularly pronounced in Japan, where demographic and social changes have intensified the risk of isolation \citep{Hirayama2021-jw, Hisata2023-sc, Sasaki2023-kj}. One key factor in this context is the rise of nuclear families, which has led to more older adults living apart from their children and grandchildren \citep{Hirayama2021-jw}, thereby reducing routine contact with family members \citep{Hisata2023-sc}. At the same time, urbanization in Japan has contributed to the weakening of local communities, which limits opportunities for older adults to interact with neighbors or community members \citep{Sasaki2023-kj}. Indeed, it has been reported that 61.1\% of older adults living alone in Japan communicate with others only once every two to three days or less \citep{Japan2024-xy}. Such limited social interaction and connection have been linked to increased risks of dementia and depression, and are recognized as major contributors to chronic anxiety and loneliness \citep{Shen2022-dk, Holt-Lunstad2024-od, Valtorta2012-lo, Nakou2025-jz}. Therefore, it is essential to provide older adults with opportunities for social interaction and meaningful connections as a form of psychological and social support.

To address this issue, dialogue agents, such as chatbots, have been increasingly developed to support older adults by providing conversational opportunities and a sense of connection \citep{An2025-rg, Chou2024-ci, Ryu2020-tb}. Unlike human counselors or caregivers, dialogue agents are not limited by human resources and can remain available at any time. Their ability to deliver immediate and continuous responses regardless of time and place enables them to offer stable opportunities for social interaction and to foster a sense of connection with the agent in daily life \citep{Mitsuno2022-ad, Klopfenstein2017-bn}. Such continuous engagement may enhance older adults' mental health \citep{Shen2022-dk, Holt-Lunstad2024-od, Valtorta2012-lo, Nakou2025-jz}. Indeed, several studies have shown that sustained interaction with dialogue agents over periods of several weeks can alleviate anxiety and depression among older adults \citep{Chou2024-ci, Ryu2020-tb}, underscoring their potential to support older adults' mental health.

However, relying solely on dialogue agents is unlikely to be sufficient to support socially isolated older adults \citep{Shen2024-ge, Liu2025-am} and could even pose long-term risks \citep{Fang2025-af}. For example, \cite{Fang2025-af} reported that when individuals with limited human connections interact exclusively with agents, they may develop excessive dependence, which increases loneliness and reduces engagement with others. In addition, the responses of dialogue agents are often perceived as more superficial and less empathetic than those of humans, suggesting that conversations with humans remain essential, particularly in situations requiring empathy, such as when sharing personal concerns \citep{Shen2024-ge, Liu2025-am}. Overall, these findings indicate that effective support for older adults requires not only interaction with agents but also mechanisms that actively foster human relationships.

\begin{figure}[t]
\centering
\includegraphics[width=1.0\linewidth]{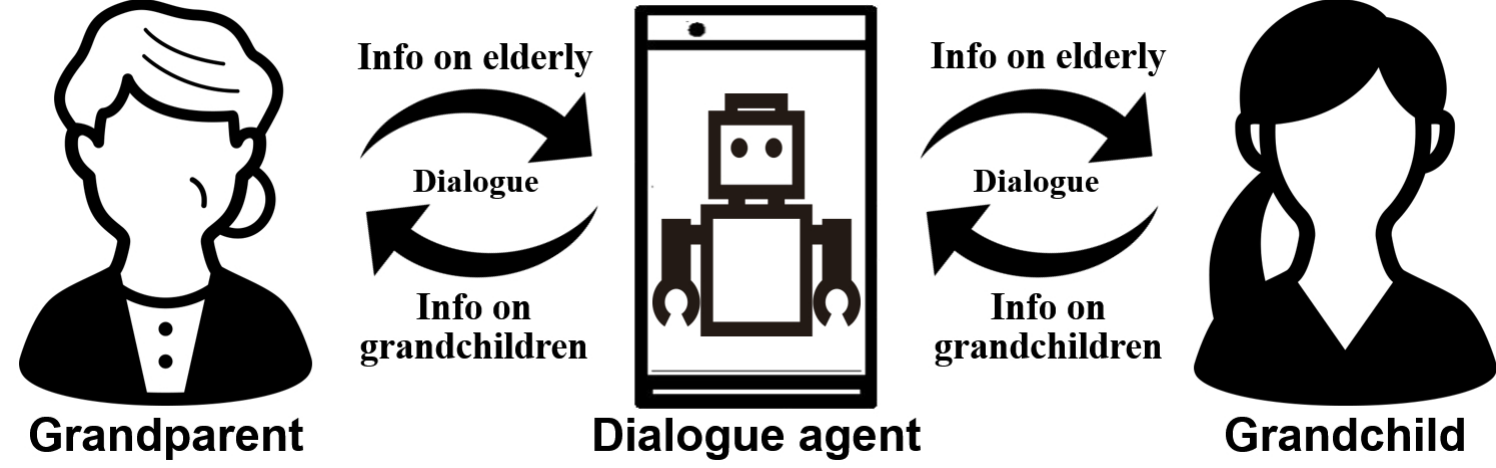}
\caption{Overview of the proposed dialogue agent. The agent serves as a conversational partner for both older adults and their grandchildren and shares information between them.}\label{fig:overview}
\end{figure}

Motivated by this need, we propose an information sharing approach in which the agent shares information obtained from one user with another during interactions. By sharing everyday information, such as daily activities and personal preferences, the agent can help users become more aware of each other's lives, thereby promoting mutual understanding and strengthening their sense of connection. In this way, information sharing may enable dialogue agents to go beyond serving as conversational partners and function as bridges that support human--human relationships. In particular, this study focuses on a dialogue agent that shares information between older adults and their grandchildren, given that prior research has shown that grandchildren represent a uniquely important source of psychological value for older adults \citep{Lai2021-ty, Yang2022-or, Gessa2020-dp}. Based on this perspective, we aim to develop a dialogue agent that supports both the relationship between older adults and the agent and that between older adults and their grandchildren.

An overview of the proposed dialogue agent is shown in Figure~\ref{fig:overview}. The agent serves as a conversational partner for both older adults and their grandchildren and shares information between them. In practice, it engages in one-on-one daily conversations separately with older adults and their grandchildren. Through these interactions, the agent collects information about each user (e.g., daily events) and periodically shares this information with the other party (e.g., ``Your grandchild mentioned that he studied with a friend at a cafe today''). By sharing such information, the dialogue agent enables older adults and their grandchildren to learn more about each other and provides opportunities for developing their relationship. As strengthened social relationships can mitigate social isolation among older adults and thereby alleviate anxiety and depressive symptoms \citep{Shen2022-dk, Holt-Lunstad2024-od, Valtorta2012-lo, Nakou2025-jz}, the proposed agent has the potential to contribute to improvements in mental health.

The remainder of this paper is organized as follows. Section~\ref{sec:hypotheses} presents our hypotheses regarding the effects of the proposed agent. Next, Section~\ref{sec:relatedwork} reviews related work on agents for supporting older adults and those that share information between users. Then, Section~\ref{sec:chatbot} describes the design of the proposed dialogue agent, followed by Section~\ref{sec:experiment}, which reports on an experiment evaluating its effectiveness. Finally, Sections~\ref{sec:discussion} and~\ref{sec:conclusion} discuss the implications of our findings and conclude the paper.

\section{Hypotheses} \label{sec:hypotheses}
We formulated the following hypotheses regarding the effectiveness of the proposed dialogue agent, which shares information between users during conversations.

\subsection{Enhancing Willingness to Interact with the Agent}
The proposed agent is designed to function as a daily conversational partner, providing users with opportunities for dialogue and a sense of connection. To achieve its intended effects, it is essential that users remain motivated to continue interacting with the agent over time. Sustained engagement enables the agent to repeatedly offer conversational opportunities and to accumulate information that can be shared between users.

Prior research suggests that people have an intrinsic interest in learning about others \citep{Way2024-ew, Renner2006-tj}, and that receiving such information can motivate continued interaction with an agent \citep{Mitsuno2022-ad}. In particular, information about close others, such as family members, is likely to be especially engaging and personally meaningful \citep{Lai2021-ty, Yang2022-or, Gessa2020-dp}. Therefore, we expect that our proposed agent, by serving not only as a conversational partner but also as a source of information about users' family members, can enhance users' motivation to interact with the agent. Based on this reasoning, we propose the following hypotheses:

\textbf{H1a:} The proposed agent will increase older adults' willingness to interact with the agent by sharing information about their grandchildren.

\textbf{H1b:} The proposed agent will increase grandchildren's willingness to interact with the agent by sharing information about their grandparents.

\subsection{Strengthening the Grandparent--Grandchild Connection}
By sharing personal information about each user with the other, the proposed agent has the potential to positively influence the relationship between older adults and their grandchildren. Prior studies have suggested that learning about another person's daily life and experiences can reduce emotional distance and promote relational closeness \citep{Fu2021-ya, Koshino2024-bl}. Accordingly, we expect that the mutual sharing of personal updates through the agent will foster a stronger sense of connection between them. Therefore, we hypothesize that:

\textbf{H2:} The proposed agent will strengthen the sense of connection between older adults and their grandchildren by sharing information about each other.

\subsection{Improving Users' Mental Health}
Finally, the proposed agent may also have beneficial effects on the mental health of both older adults and grandchildren. By sharing personal information between them, the agent is expected to increase users' willingness to engage in conversations with the agent (H1) and to strengthen the psychological connection between older adults and their grandchildren (H2). As increased opportunities for social interaction and stronger interpersonal connections are associated with reduced anxiety and depressive symptoms \citep{Shen2022-dk, Holt-Lunstad2024-od, Valtorta2012-lo, Nakou2025-jz}, these effects are likely to contribute to improved mental health outcomes. Based on this perspective, we propose the following hypotheses:

\textbf{H3a:} The proposed agent will contribute to improving older adults' mental health by sharing information about their grandchildren.

\textbf{H3b:} The proposed agent will contribute to improving grandchildren's mental health by sharing information about their grandparents.

\section{Related Work} \label{sec:relatedwork}
\subsection{Dialogue Agents for Supporting Older Adults}
As population aging progresses \citep{Xi2025-rn}, increasing attention has been directed toward interactive technologies aimed at supporting older adults \citep{Wada2006-zj, Tan2024-ej, Broekens2009-in}. In particular, many studies have focused on developing dialogue agents that provide conversational opportunities for older adults \citep{Abdollahi2017-jc, Nishio2021-lv, Ryu2020-tb}. For example, Abdollahi et al. \citep{Abdollahi2017-jc} proposed Ryan, a companion robot designed for older adults, and showed that interacting with the robot on a daily basis improved users' mental health. Similarly, \cite{Nishio2021-lv} developed a system in which two dialogue robots collaboratively engaged with older adults, showing that this design increased self-disclosure and contributed to improvements in users' emotional states. Furthermore, with the growing prevalence of smartphones among older adults \citep{MIC2023-xy}, chatbots have also been explored as accessible tools for providing conversational support to older adults. For instance, \cite{Ryu2020-tb} developed the Mental Health Care Chatbot (MHCC), which provided daily conversations for older adults and was found to reduce depressive tendencies and anxiety. Together, these studies suggest that dialogue agents can serve as effective tools for offering conversational opportunities, thereby supporting the mental health of older adults.

However, most previous studies have primarily focused on one-to-one interactions between the agent and older adults. Such a design may lead to unintended consequences, particularly for socially isolated older adults, as prolonged reliance on agents as sole conversational partners can result in overdependence, reduced human interaction, and heightened feelings of loneliness \citep{Fang2025-af}. To address this issue, the present study aims to develop a dialogue agent that not only serves as a daily conversational partner for older adults but also facilitates communication and connection with their grandchildren. The proposed agent is designed to foster two types of relationships: the human--agent relationship between the older adult and the agent, and the human--human relationship between the older adult and the grandchild. Through this dual design, the agent is expected to enhance conversational opportunities and promote social connectedness through both human--agent and human--human interactions.

\subsection{Dialogue Agents as Information Sharing Partners}
The present study aims to develop a dialogue agent that simultaneously promotes relationship building between older adults and the agent, as well as between older adults and their grandchildren. To support this dual relationship formation, we focus on a dialogue strategy in which the agent shares information about other users during conversations. Prior research in social psychology has shown that learning about others increases interest, familiarity, and intimacy, which in turn fosters the development of closer interpersonal relationships \citep{Altman1973-hv, Collins1994-tp, Guerrero2017-vk}. Building on these insights, a few HAI studies have explored the use of dialogue agents that share personal information to support interpersonal connectedness \citep{Fu2021-ya, Koshino2024-bl}. These studies suggest that sharing personal information through an agent may help foster interpersonal relationships.

While prior work has demonstrated the potential of such approaches, it has primarily focused on short-term interactions in controlled laboratory environments. In contrast, the present study develops and empirically evaluates a dialogue agent for everyday use, with an emphasis on long-term interaction and practical deployment. Specifically, we focus on the following three aspects.

\textbf{1. Use of a chatbot platform.}
Previous studies mainly employed physically embodied robots \citep{Fu2021-ya, Koshino2024-bl}. In this study, we adopt a chatbot as an alternative dialogue platform to enable more practical and scalable deployment. A smartphone-based chatbot allows users to engage in conversations anytime and anywhere at minimal cost, making it well suited for large-scale, real-world use \citep{Mitsuno2022-ad, Klopfenstein2017-bn}. If this chatbot-based approach effectively supports users through information sharing, it would demonstrate the potential of chatbots as an accessible and widely deployable solution. Moreover, smartphone use among older adults has increased substantially, with 90.2\% of those in their 60s and 67.0\% of those in their 70s using smartphones in Japan \citep{MIC2023-xy}. This trend suggests that chatbot-based approaches are increasingly feasible for supporting older adults in real-world settings. Based on this context, we develop a chatbot designed to share personal information and support social connection between users.

\textbf{2. Long-term dialogue design.}
In contrast to prior work that focused on brief interactions (e.g., a few sessions lasting several minutes) \citep{Fu2021-ya, Koshino2024-bl}, this study aims to enable longer-term interactions between users and the agent. Supporting long-term engagement is important because it allows the agent to provide continuous conversational opportunities and to accumulate personal information that can be shared over time, thereby enhancing both user--agent interaction and interpersonal connection. To achieve this, we design an agent capable of engaging in daily conversations over an extended period (e.g., ten days). Specifically, we implement agent-initiated dialogues that adapt to each user's lifestyle. The agent estimates each user's wake-up time and bedtime and automatically determines when to send messages, promoting intermittent interactions throughout the day (e.g., around five times per day). This intermittent yet agent-driven approach enables users to maintain interaction with low cognitive effort, facilitating long-term engagement.

\textbf{3. Evaluation with older adults.}
Previous studies have primarily focused on young adults such as university students \citep{Fu2021-ya, Koshino2024-bl}. In this study, we instead focus on older adults as a key population for evaluating real-world impact of information-sharing agents. In Japan, population aging is rapidly progressing \citep{Xi2025-rn}, and nuclear families have become increasingly common \citep{Hirayama2021-jw}. As a result, older adults are at increased risk of social isolation and represent an important group that may benefit from these technologies. However, few studies have explored the use of chatbots by older adults in naturalistic settings. To address this gap, we conduct a ten-day experiment involving older adults to examine the effectiveness of the proposed agent and to provide empirical evidence on its feasibility and potential psychological benefits.

In summary, the present study advances the design of dialogue agents that share personal information in three key ways: (1) implementing such agents on a practically deployable chatbot platform, (2) enabling sustained, long-term interactions through an agent-initiated, intermittent dialogue design that adapts to users' daily routines, and (3) empirically evaluating the effectiveness of the proposed agent with older adults. This approach allows us to examine the potential of these agents in realistic and socially relevant contexts.

\section{Chatbot} \label{sec:chatbot}

\begin{table*}[tbp]
	\begin{center}
		\caption{Daily Dialogue flow.}
		\label{table:daily_dialog_flow}
		\begin{tabular}[hbt]{l l l l}
			\hline
			\bf Context & \bf Timing &  \bf Topic & \bf Example\\
			\hline
			Morning    & 0.5 hours after waking up & Sleep	&  What time did you go to bed last night? \\
			Noon       & 4 hours after waking up & Meal	&  What did you eat for lunch?	\\
			Afternoon  & 6 hours after waking up & Location    &  Where are you now? \\
			Evening      & 4 hours before sleeping & Impression	&  How was your day today?	\\
			Night & 2 hours before sleeping & Plan	&  What's your schedule for tomorrow? \\
			-          & Unspecified$^{*1}$	& Value	&  What is your favorite season? 	\\
			\hline
		\end{tabular}
	\end{center}
    
    \begin{threeparttable}[h]
        \begin{tablenotes}
            \item[*1] Questions about Value are asked with a 40\% probability, replacing one of the other question types (Sleep, Meal, Location, Impression, or Plan).
        \end{tablenotes}
    \end{threeparttable}
\end{table*}

\subsection{System Architecture}
\begin{figure}[t]
\centering
\includegraphics[width=1.0\linewidth]{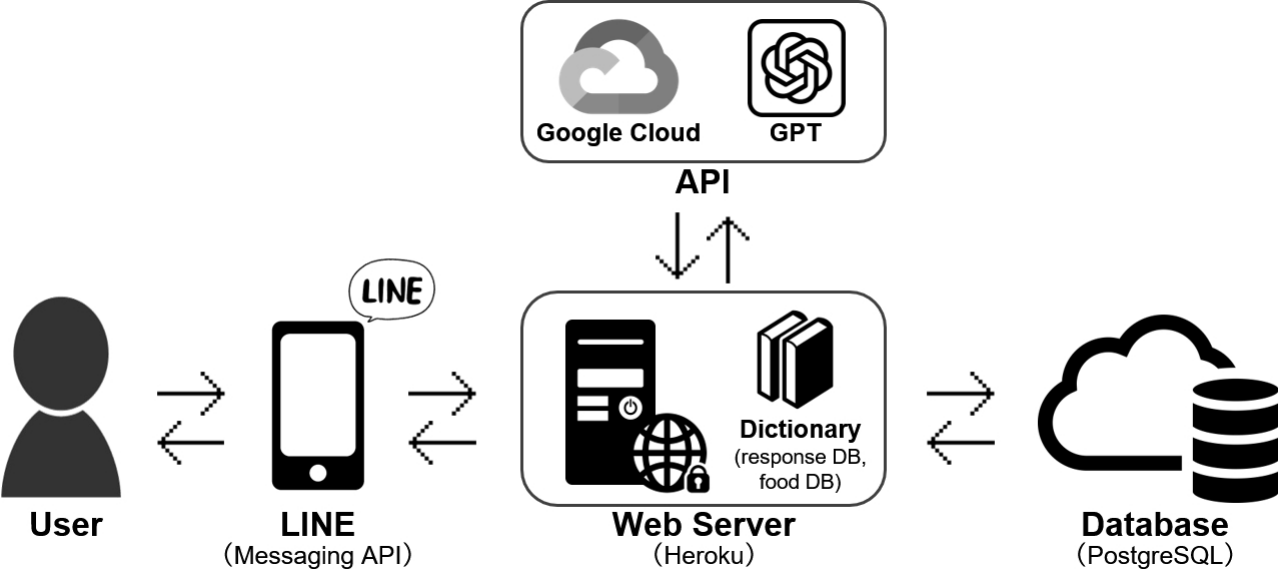}
\caption{System architecture.}\label{fig:system_architecture}
\end{figure}

Figure~\ref{fig:system_architecture} illustrates the system architecture of the proposed chatbot. The chatbot operates on LINE, Japan's most widely used messaging application \citep{Steinberg2020-tx}, and communicates with users via the LINE Messaging API. It utilizes a dataset that pairs anticipated user utterances with corresponding agent responses, along with a list of food names constructed from Cookpad's data \citep{Cookpad2015-aa}. Additionally, the system employs the Google Natural Language API for sentiment and entity analysis, the OpenAI API for generating contextually appropriate responses, and a SQL database for storing and retrieving users' dialogue histories.

\begin{table*}[tbp]
	\begin{center}
		\caption{An example dialogue between the older adult and the agent, taken from an actual experimental session with the topic ``Location.'' $\langle grandchild\ name\rangle$ denotes the name of the user's grandchild.}
		\label{table:dialog_example}
		\begin{tabular}[hbt]{p{0.05\linewidth} p{0.1\linewidth} p{0.7\linewidth}}
			\hline
			\bf Turn & \bf Speaker & \bf Utterance\\
			\hline
			1 & Agent    & Hey, where are you spending time right now?  \\
			2 & User     & I'm at home. I was cleaning the shoe rack. \\
			3 & Agent    & Oh, I see. I heard that $\langle grandchild\ name\rangle$ was also spending some time at home the other day. Home really does sound like a comfortable place. \\
			4 & User     & Definitely. It's nice and relaxing, and it's cool in here. \\
			5 & Agent    & Yeah, even on hot days, it feels nice and cozy when you're at home. \\
			\hline
		\end{tabular}
	\end{center}
\end{table*}

\subsection{Dialogue Design} \label{sec:dialogue_design}
When designing the agent's dialogues, we incorporate several features to facilitate long-term interaction. First, we implement an agent-initiated dialogue strategy, in which the agent proactively asks the user questions. This allows users to engage in conversations without having to initiate them spontaneously. Such a design helps prevent users from discontinuing interactions after a short period of time due to difficulty in coming up with conversational topics.

Next, we design the dialogues to be intermittent and synchronized with the user's daily routine (Table~\ref{table:daily_dialog_flow}). For example, the agent asks about users' sleep in the morning, their activities during the day, and their impressions of the day in the evening. The timing of each question is determined based on the user's estimated wake-up time and bedtime, which are inferred from previous responses to sleep-related questions. If a user does not respond by the next scheduled question, the agent skips the unanswered question and proceeds. However, if no response is received for more than 20 hours, the agent assumes that the user will not reply and generates the next question. Additionally, if no response is received within six hours after a question is generated, the agent sends a gentle reminder to encourage a reply.

Furthermore, the dialogue between the user and the agent is designed to resemble a simple SMS-style exchange with a friend. Each conversation is short and casual, lasting approximately one minute, and covers light topics such as daily activities or personal preferences. Each dialogue sequence consists of five basic turns (Table~\ref{table:dialog_example}):

\begin{enumerate}[label=(\arabic*)]
    \item the agent's question,
    \item the user's answer,
    \item the agent's comment on the answer,
    \item the user's response to the comment, and
    \item the agent's reply, which could be either a text message or a sticker.
\end{enumerate}

The agent generates its utterances in turns 1, 3, and 5 using a combination of predefined rules and a large language model (LLM). Specifically, in turn 1, the agent selects a question from a predefined set prepared for each dialogue topic (Table~\ref{table:dialog_example}). In turn 3, the agent's comment on the user's answer is generated either by rule-based template matching or by the LLM (details are described in Section~\ref{sec:sharing_info}). Finally, in turn 5, the agent's reply is determined based on the content and length of the user's message in turn 4. When the user's message contains five or fewer characters and matches a predefined response (e.g., ``Thank you''), the agent sends the corresponding reply (e.g., ``You're welcome''). If the short message does not match any predefined response, the agent sends a sticker instead. For longer utterances (more than five characters), the agent generates an appropriate text response using the LLM. When the user sends an additional message beyond turn 5, the agent no longer generates text responses using the LLM. Instead, it applies a simplified rule: if the new message matches a predefined response, the agent replies with the corresponding phrase; otherwise, it sends a sticker.

\subsection{Sharing Others' Information} \label{sec:sharing_info}
\begin{figure}[t]
\centering
\includegraphics[width=0.95\linewidth]{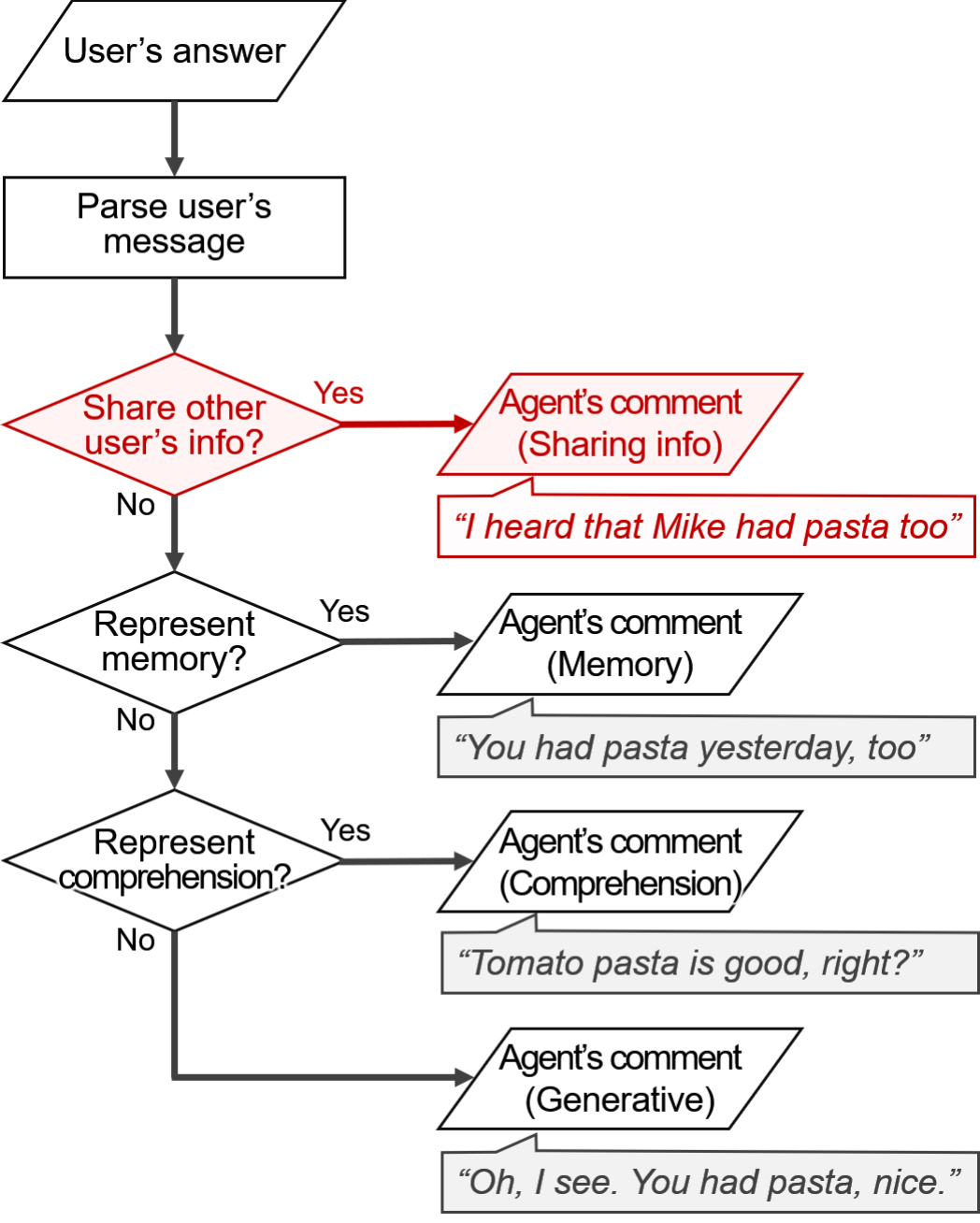}
    \caption{Decision flow of the agent's comment generation. The red components appear only in the proposed Sharing condition, while the baseline Non-sharing condition always follows the ``No'' path at this stage.}
    \label{fig:comment_flow}
\end{figure}

A key feature of the proposed agent is its ability to share information about other users during conversations. The agent produces such statements in turn 3. Examples include ``I heard that Mike also had pasta,'' ``I heard that Mike is in Tokyo today,'' and ``I heard that Mike likes autumn.'' To prevent the agent's comments from becoming repetitive, these information-sharing statements are generated only under specific conditions. Figure~\ref{fig:comment_flow} illustrates the decision flow used to determine the agent's comment in turn 3.

The agent's comment in turn 3 could take one of four forms: (A) a response that shares information about other users (\textit{Sharing info}), (B) a response generated from the agent's memory of the user's past statements (\textit{Memory}), (C) a response based on the agent's understanding of the content (\textit{Comprehension}), or (D) a response generated by a language model (\textit{Generative}). These response types are evaluated sequentially in the following order: \textit{Sharing info}, \textit{Memory}, \textit{Comprehension}, and finally \textit{Generative}, until one applicable response type is found (Figure~\ref{fig:comment_flow}).

A \textit{Sharing info} response is generated only when two conditions are met:
(A) the agent possesses relevant information about the other participant, and
(B) a 40\% probability condition is satisfied. As an exception, this probability is increased to 80\% for conversations on the topic of ``Value,'' as these are considered more conducive to interpersonal connection. A \textit{Memory} response is produced when the user has previously made a similar statement. A \textit{Comprehension} response is generated when the system successfully recognizes specific entities in the user's utterance (e.g., food names or place names). Both of these responses are created using rule-based template matching. When none of the above conditions is satisfied, the agent generates a \textit{Generative} response using the LLM.

\section{Experiment} \label{sec:experiment}
\subsection{Experimental Design}
The experiment was conducted using a between-subjects design with two conditions: the Non-sharing and Sharing conditions. The Non-sharing condition represents a conventional chatbot designed to support interaction solely between a user and the agent, in which the chatbot engages in conversations with both older adults and their grandchildren but does not share any personal information between them. In contrast, the Sharing condition represents the proposed chatbot, which extends conventional human--agent interaction by sharing personal information about older adults and their grandchildren during their respective conversations. More specifically, the \textit{Sharing info} response was generated only in the Sharing condition, while it was always omitted in the Non-sharing condition (see Figure~\ref{fig:comment_flow}).

\subsection{Participants}
A total of 104 participants (52 pairs) took part in the experiment, each pair consisting of one older adult and one grandchild. The older adults (10 men, 42 women) had a mean age of 77.44 years ($SD = 4.00$), while the grandchildren (25 men, 27 women) had a mean age of 22.25 years ($SD = 4.32$). All participants were native Japanese speakers living in Japan.

Participants were required to own a smartphone, be able to use LINE, a widely used messaging application in Japan \citep{Steinberg2020-tx}, and live separately from their paired partner. In addition, older adults had to be 65 years or older, and grandchildren had to be 18 years or older.

\subsection{Procedure}
This experiment was approved by the Ethics Committee for Research Involving Human Subjects at the Graduate School of Engineering Science, Osaka University. All participants provided informed consent prior to participation. In the Sharing condition, participants were informed that portions of their conversation with the chatbot might be shared with the other user (their partner).

Before starting the interaction with the chatbot, participants completed a pre-questionnaire. They then configured their smartphones to enable interaction with the chatbot via the LINE application. During this setup, each participant registered their name as well as their typical wake-up time and bedtime for both weekdays and weekends. Participants were instructed as follows: ``You will receive several messages from the chatbot each day. Please reply to them. There is no specific time requirement for your responses, but try to reply whenever possible.'' Following this instruction, participants interacted with the chatbot for ten consecutive days. After completing the ten-day interaction period, they completed a post-questionnaire.

\subsection{Measurements}
To examine Hypothesis~1, participants' willingness to interact with the agent was assessed. In the post-questionnaire, participants rated their willingness to engage with the agent using the Intention to Use scale \citep{Heerink2010-cl}, which is widely used to evaluate motivation for engaging with dialogue agents \citep{Mahzoon2019-la, Mitsuno2024-tb}. In addition to this self-report measure, behavioral indicators of engagement were analyzed, as actual interaction patterns provide valuable complementary evidence of subjective motivation. Following \cite{Mitsuno2022-ad}, who evaluated willingness to interact with a chatbot, two behavioral measures were adopted: (1) Average Response Time, defined as the mean time interval between the chatbot's message and the user's reply, and (2) Number of Reminders, defined as the number of times the chatbot sent a follow-up message six hours after receiving no response.

To test Hypothesis~2, changes in perceived connection between older adults and their grandchildren were assessed. In both the pre- and post-questionnaires, participants rated perceived psychological closeness with their partner using the Inclusion of Other in the Self (IOS) scale \citep{Aron1992-jm}. Furthermore, participants provided open-ended responses in the post-questionnaire, describing how their connection with their partner (the grandparent or grandchild) had changed through interactions with the chatbot. These qualitative responses are particularly important for capturing subtle or individualized effects that may not emerge in aggregated quantitative scores.

To evaluate Hypothesis~3, participants' mental health was assessed using the Hospital Anxiety and Depression (HAD) scale \citep{Zigmond1983-sr}, administered in both the pre- and post-questionnaires. This scale consists of two subscales: anxiety and depression. Furthermore, in the post-questionnaire, participants provided open-ended responses describing any perceived changes in their mental state resulting from their interactions with the chatbot. These responses capture subjective changes in mental health beyond what is reflected in scale-based measures.

To detect satisficing behavior, in which participants respond carelessly or without sufficient attention \citep{Oppenheimer2009-sl}, an item based on the Directed Questions Scale (DQS) \citep{Maniaci2014-hr} was included in the post-questionnaire. Specifically, participants were instructed to select a particular response option (``Please choose 1 for this item'') to verify their attentiveness.

\subsection{Results}
\subsubsection{Frequency of \textit{Sharing info} Responses}
In the Sharing condition, the chatbot selected a \textit{Sharing info} response in 28.85\% of all turn-3 comment opportunities. On average, participants received 11.60 such responses ($SD = 3.88$), with counts ranging from 4 to 20. These results confirm that all participants in the Sharing condition were exposed to multiple \textit{Sharing info} responses over the ten-day interaction period.

\subsubsection{Data Screening and Coding}
\paragraph{Screening}
Among the 104 participants, 23 older adults and 5 grandchildren failed the DQS item. Prior research in survey methodology suggests that older adults are more susceptible to satisficing behavior due to increased cognitive load and reduced attentional resources \citep{Schneider2022-hd}. Accordingly, analyses of subjective measures (Intention to Use, IOS, and HAD) included only participants who passed the DQS check. 

Nevertheless, since our inspection of open-ended responses and chatbot interaction logs suggested that participants who failed the DQS item still engaged with the experiment in a meaningful and consistent manner, their data were retained for subsequent analyses. Specifically, for these participants, the open-ended responses regarding perceived connection averaged 10.79 characters in Japanese ($SD = 6.12$), with lengths ranging from 2 to 23 characters; all responses were semantically interpretable. Similarly, their open-ended responses about mental health averaged 16.71 characters ($SD = 12.62$), ranging from 2 to 50 characters, and all statements conveyed meaningful content. Notably, the shortest two-character response was ``no change.'' Additionally, an examination of chatbot interaction logs confirmed that these participants actively engaged with the chatbot. The number of messages they sent during the ten-day period averaged 118.82 ($SD = 30.15$), ranging from 65 to 205, indicating that they participated consistently in the daily conversations. Therefore, their data were included in the behavioral analyses of engagement (Average Response Time and Number of Reminders).

\paragraph{Coding of Open-Ended Responses}
Next, to enable quantitative analysis of the open-ended responses, we conducted a coding procedure. Two sets of responses, those concerning perceived connection and those concerning mental health, were coded by independent third-party coders who were blind to the study purpose and experimental condition. Each response was categorized into one of three groups: positive change, no change, or negative change.

For each measure, 20 coders (10 men and 10 women) performed the classification. The coders' mean age was 41.70 years ($SD = 7.66$) for the perceived-connection responses and 40.90 years ($SD = 10.05$) for the mental-health responses. Fleiss's $\kappa$ coefficients indicated high inter-rater agreement for both perceived connection ($\kappa = .759$, 95\% CI [.748, .771], $p < .001$) and mental health ($\kappa = .893$, 95\% CI [.881, .905], $p < .001$).

Following prior research on coding open-ended responses \citep{Tornberg2025-rt}, the final label assigned to each participant's response was determined by the category most frequently selected among the 20 coders. In the few instances where the top categories were tied (one case per measure), the authors discussed the responses and reached an agreement on the final label.

\subsubsection{Hypothesis Testing}
\paragraph{Willingness to Interact with the Agent (H1)}
We first examined the subjective measure assessed in the post-questionnaire, Intention to Use, by conducting Welch's $t$ tests separately for older adults and grandchildren. For older adults, there was no significant difference between the Sharing condition ($M = 3.75$, $SD = 1.13$) and the Non-sharing condition ($M = 3.43$, $SD = 1.43$) ($t(26.59) = 0.669$, $p = .509$, $d = 0.24$). Similarly, for grandchildren, no significant difference was observed between the Sharing condition ($M = 2.71$, $SD = 1.39$) and the Non-sharing condition ($M = 2.36$, $SD = 1.51$) ($t(44.35) = 0.816$, $p = .419$, $d = 0.24$).

\begin{figure}[t]
\centering
\includegraphics[width=1.0\linewidth]{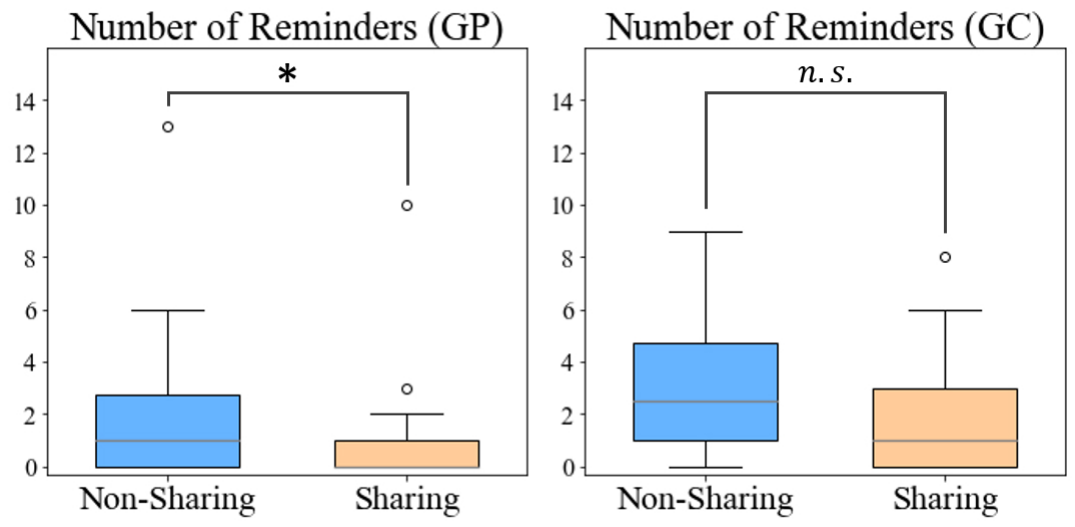}
\caption{Number of reminders sent to users ($^{*}p < .05$). A reminder was issued when a user did not reply to the chatbot for more than six hours. GP = grandparent; GC = grandchild.}
\label{fig:reminder}
\end{figure}

Next, we analyzed the behavioral measures, Average Response Time and Number of Reminders, using Mann--Whitney $U$ tests for older adults and grandchildren separately (Figure~\ref{fig:reminder}). For Average Response Time, among older adults, there was no significant difference between the Sharing condition ($\mathrm{Mdn} = 34.41$ min, $\mathrm{IQR} = [20.70,\,63.74]$) and the Non-sharing condition ($\mathrm{Mdn} = 48.64$ min, $\mathrm{IQR} = [33.75,\,73.19]$) ($U = 274$, $p = .241$, $r = .16$). Similarly, for grandchildren, no significant difference was found between the Sharing condition ($\mathrm{Mdn} = 67.70$ min, $\mathrm{IQR} = [39.57,\,107.93]$) and the Non-sharing condition ($\mathrm{Mdn} = 76.03$ min, $\mathrm{IQR} = [48.81,\,122.95]$) ($U = 289$, $p = .370$, $r = .12$).

For Number of Reminders, older adults in the Sharing condition ($\mathrm{Mdn} = 0.00$, $\mathrm{IQR} = [0.00,\,1.00]$) received significantly fewer reminders than those in the Non-sharing condition ($\mathrm{Mdn} = 1.00$, $\mathrm{IQR} = [0.00,\,2.75]$) ($U = 212$, $p = .013$, $r = .35$). In contrast, for grandchildren, no significant difference was observed between the Sharing condition ($\mathrm{Mdn} = 1.00$, $\mathrm{IQR} = [0.00,\,3.00]$) and the Non-sharing condition ($\mathrm{Mdn} = 2.50$, $\mathrm{IQR} = [1.00,\,4.75]$) ($U = 274$, $p = .241$, $r = .16$). Taken together, these results suggest that older adults in the Sharing condition were less likely to leave the chatbot's questions unanswered for extended periods, indicating a higher level of behavioral engagement with the agent.

\begin{table*}[tbp]
    \begin{center}
        \caption{Coded results of changes in perceived connection. Numbers in parentheses indicate adjusted residuals, with asterisks denoting significance levels ($|r| \ge 1.95$: $^{*}$, $|r| \ge 2.58$: $^{**}$).}
        \label{table:perceived_connection}
        \begin{tabular}[hbt]{llccc}
            \hline
             & \bf Condition & \bf Positive & \bf No change & \bf Negative \\
            \hline
            \multirow{2}{*}{Older adults}
                & Sharing      & 14 (2.91$^{**}$)  & 12 ($-2.59$$^{**}$) & 0 (1.01) \\
                & Non-sharing  & 4 ($-2.91$$^{**}$) & 21 (2.59$^{**}$)   & 1 ($-1.01$) \\
            \hline
            \multirow{2}{*}{Grandchildren}
                & Sharing      & 16 (2.81$^{**}$)  & 10 ($-2.51$$^{*}$) & 0 ($-1.01$) \\
                & Non-sharing  & 6 ($-2.81$$^{**}$) & 19 (2.51$^{*}$)   & 1 (1.01) \\
            \hline
        \end{tabular}
    \end{center}
\end{table*}

\paragraph{Grandparent--Grandchild Connection (H2)}
For the IOS scores assessed in the pre- and post-questionnaires, we conducted 2 (Condition: Sharing vs.\ Non-sharing) $\times$ 2 (Time: Pre vs.\ Post) mixed ANOVAs separately for older adults and grandchildren. For older adults, neither the main effect of Condition ($F(1, 27) = 0.004$, $p = .951$, $\eta^2 = .000$) nor the main effect of Time ($F(1, 27) = 1.810$, $p = .190$, $\eta^2 = .063$) was significant. The interaction effect was also not significant ($F(1, 27) = 0.565$, $p = .459$, $\eta^2 = .020$). Likewise, for grandchildren, neither the main effect of Condition ($F(1, 45) = 0.309$, $p = .581$, $\eta^2 = .007$) nor the main effect of Time ($F(1, 45) = 0.565$, $p = .456$, $\eta^2 = .012$) was significant, and the interaction was also not significant ($F(1, 45) = 1.319$, $p = .257$, $\eta^2 = .028$).

Next, we examined changes in perceived connection based on the coded open-ended responses and conducted chi-square tests separately for older adults and grandchildren (Table~\ref{table:perceived_connection}). For older adults, a significant difference was observed between the Sharing condition (positive: 14, no change: 12, negative: 0) and the Non-sharing condition (positive: 4, no change: 21, negative: 1) ($\chi^2(2) = 9.010$, $p = .011$, $V = .42$). Residual analysis showed that positive responses were significantly more frequent in the Sharing condition (adjusted residual = 2.91), whereas no-change responses were significantly more frequent in the Non-sharing condition (adjusted residual = 2.59). A similar pattern was observed for grandchildren: a significant difference emerged between the Sharing condition (positive: 16, no change: 10, negative: 0) and the Non-sharing condition (positive: 6, no change: 19, negative: 1) ($\chi^2(2) = 8.339$, $p = .015$, $V = .40$). Residual analysis again indicated significantly more positive responses in the Sharing condition (adjusted residual = 2.81) and significantly more no-change responses in the Non-sharing condition (adjusted residual = 2.51). These results suggest that participants in the Sharing condition perceived positive improvements in their connection with their partner through interactions with the chatbot.

\begin{figure}[t]
\centering
\includegraphics[width=1.0\linewidth]{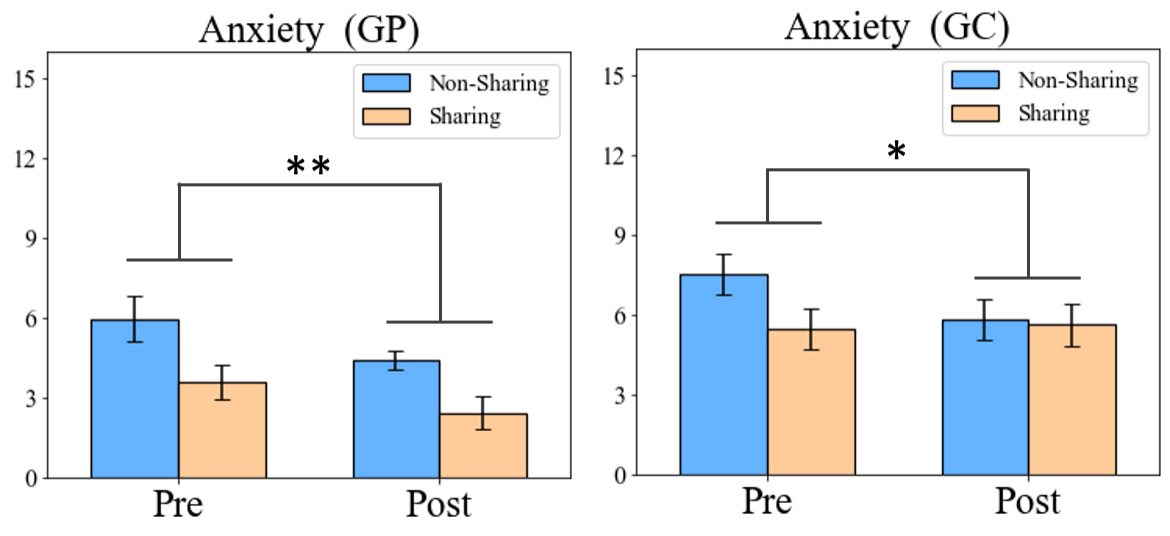}
\caption{Anxiety scores before and after the interaction with the chatbot ($^{*}p < .05, ^{**}p < .01$). Error bars represent standard errors. GP = grandparent; GC = grandchild.}
\label{fig:anxiety}
\end{figure}

\paragraph{Users' Mental Health (H3)}
For the HAD scores assessed in the pre- and post-questionnaires, we conducted 2 (Condition: Sharing vs.\ Non-sharing) $\times$ 2 (Time: Pre vs.\ Post) mixed ANOVAs separately for older adults and grandchildren (Figure~\ref{fig:anxiety}). For Anxiety, older adults showed a significant main effect of Time ($F(1, 27) = 10.529$, $p = .003$, $\eta^2 = .281$), suggesting a decrease in anxiety over the course of interacting with the chatbot. In contrast, neither the main effect of Condition ($F(1, 27) = 3.002$, $p = .095$, $\eta^2 = .100$) nor the interaction effect ($F(1, 27) = 0.069$, $p = .794$, $\eta^2 = .003$) was significant. For grandchildren, the main effect of Time was also significant ($F(1, 45) = 5.715$, $p = .021$, $\eta^2 = .113$), again suggesting a reduction in anxiety following the interaction with the chatbot. However, neither the main effect of Condition ($F(1, 45) = 0.594$, $p = .445$, $\eta^2 = .013$) nor the interaction ($F(1, 45) = 3.796$, $p = .058$, $\eta^2 = .078$) was significant. 

For Depression, older adults did not show a significant main effect of Condition ($F(1, 27) = 0.215$, $p = .647$, $\eta^2 = .008$), Time ($F(1, 27) = 0.165$, $p = .688$, $\eta^2 = .006$), or interaction ($F(1, 27) = 1.762$, $p = .196$, $\eta^2 = .061$). Similarly, for grandchildren, no significant main effect of Condition ($F(1, 45) = 0.371$, $p = .546$, $\eta^2 = .008$), Time ($F(1, 45) = 0.000$, $p = .987$, $\eta^2 = .000$), or interaction ($F(1, 45) = 0.631$, $p = .431$, $\eta^2 = .014$) was found.

\begin{table*}[tbp]
    \begin{center}
        \caption{Coded results of changes in mental health. Numbers in parentheses indicate adjusted residuals.}
        \label{table:mental_health}
        \begin{tabular}[hbt]{llccc}
            \hline
             & \bf Condition & \bf Positive & \bf No change & \bf Negative \\
            \hline
            \multirow{2}{*}{Older adults}
                & Sharing      & 17 (0.00) & 8 (0.00)  & 1 (0.00) \\
                & Non-sharing  & 17 (0.00) & 8 (0.00)  & 1 (0.00) \\
            \hline
            \multirow{2}{*}{Grandchildren}
                & Sharing      & 12 ($-0.28$) & 13 (0.56)  & 1 ($-0.59$) \\
                & Non-sharing  & 13 (0.28)   & 11 ($-0.56$) & 2 (0.59) \\
            \hline
        \end{tabular}
    \end{center}
\end{table*}

Next, we analyzed changes in mental health reflected in the coded open-ended responses and conducted chi-square tests separately for older adults and grandchildren (Table~\ref{table:mental_health}). For older adults, no significant difference was found between the Sharing condition (positive: 17, no change: 8, negative: 1) and the Non-sharing condition (positive: 17, no change: 8, negative: 1) ($\chi^2(2) = 0.000$, $p = 1.000$, $V = .00$). Similarly, for grandchildren, no significant difference was observed between the Sharing condition (positive: 12, no change: 13, negative: 1) and the Non-sharing condition (positive: 13, no change: 11, negative: 2) ($\chi^2(2) = 0.540$, $p = .763$, $V = .10$).

\section{Discussion} \label{sec:discussion}
\subsection{Willingness to Interact with the Agent (H1)}
Hypothesis~1 proposed that users' willingness to interact with the agent would increase when the agent shared information about either the older adult or the grandchild during conversations. The experimental results showed that, among older adults, the behavioral measure Number of Reminders was significantly lower in the Sharing condition than in the Non-sharing condition. Given that reminders were sent when a user did not respond to the agent's question for more than six hours, this result suggests that older adults were more motivated to engage with an agent that provided updates about their grandchild, leading to more active and consistent responses. This finding provides partial support for H1a, particularly from a behavioral perspective.

One plausible explanation for this increased engagement lies in the unique psychological value that grandchildren hold for older adults. Prior studies have shown that grandchildren provide older adults with a strong sense of purpose, meaning, and generativity \citep{Lai2021-ty, Yang2022-or, Gessa2020-dp}. In light of this, conversations in which the agent shared everyday information about the grandchild may have been perceived as more attractive and enjoyable, making interactions with the agent a more meaningful part of daily life. This increased willingness to interact with the agent is particularly important because it is essential for maintaining long-term engagement and for fostering a relationship between older adults and the agent \citep{Mahzoon2019-la,Mitsuno2024-tb}, both of which are key factors in enabling effective conversational assistance for older adults.

In contrast, no significant differences were observed between conditions in any of the subjective or behavioral measures for grandchildren, and thus H1b was not supported. One possible reason for this contrast between older adults and grandchildren is that they differ in the degree to which they are interested in each other's daily lives. Although both groups may regard one another as important family members, prior work suggests that older adults tend to ascribe particularly strong psychological significance to their grandchildren, extending beyond simple affection and relating to a sense of generational continuity \citep{Lai2021-ty}. Given this asymmetry, the types of information shared by the agent in this study, such as meals, sleep times, and current locations, may have been sufficiently engaging for older adults but not inherently compelling for grandchildren. Future research should therefore examine how to infer, collect, and share information about older adults that is likely to be more meaningful and appealing to grandchildren, thereby enabling reciprocal motivation to interact with the agent.

\subsection{Grandparent--Grandchild Connection (H2)}
We next examined Hypothesis~2, which proposed that sharing personal information through the agent would strengthen the connection between older adults and their grandchildren. The results provided partial support for this hypothesis. While no significant differences were observed between conditions in the IOS scale, the coded open-ended responses showed that participants in the Sharing condition reported positive changes in perceived connection significantly more often than those in the Non-sharing condition. To illustrate the kinds of positive changes observed in the Sharing condition, we present several representative comments. Older adults wrote, for example, ``I felt closer to my grandchild and came to understand more about their daily life'' and ``It made me want to see them.'' Grandchildren expressed similar sentiments, stating ``I learned unexpected things about my grandmother and discovered shared similarities, which made me feel closer to her'' and ``I feel more connected.'' These responses suggest that sharing everyday updates helped the two better understand each other, which may in turn strengthened their psychological closeness. Social psychological research has shown that exposure to personal information about others increases interest, familiarity, and feelings of closeness, ultimately promoting the development of interpersonal relationships \citep{Altman1973-hv, Collins1994-tp, Guerrero2017-vk}. The observed enhancement in perceived connection aligns with this theoretical background and indicates that even seemingly minor details from daily life may help individuals feel closer to one another.

\subsection{Users' Mental Health (H3)}
The proposed agent was designed to increase users' conversational opportunities and support the development of social connectedness through two mechanisms: promoting dialogue between users and the agent (H1), and strengthening the psychological connection between older adults and their grandchildren (H2). These increased conversational opportunities and enhanced social connectedness may contribute to improvements in users' mental health, particularly by reducing anxiety and depressive symptoms (H3) \citep{Shen2022-dk, Holt-Lunstad2024-od, Valtorta2012-lo, Nakou2025-jz}.

The results showed that both older adults and grandchildren exhibited significant decreases in Anxiety scores from pre- to post-interaction in both the Sharing and Non-sharing conditions. In the open-ended responses as well, participants in both conditions frequently reported positive changes in their mental state, and negative changes were almost never mentioned (with zero negative responses in the Sharing condition), although the overall distribution of responses did not differ significantly between conditions. Representative comments included: ``Although it was brief, I felt enjoyment and a sense of fulfillment in talking every day'' (older adult, Non-sharing); ``It was enjoyable and fulfilling'' (older adult, Sharing); ``Through conversations with the bot, I was able to ease my usual feelings of loneliness'' (grandchild, Non-sharing); and ``I felt emotionally enriched'' (grandchild, Sharing). These responses suggest that interacting with the agent, regardless of condition, provided a form of emotional support that may have contributed to improved mental health.

Although users' mental health improved overall, the magnitude of improvement did not differ significantly between conditions, offering only partial support for H3, which predicted greater benefits in the Sharing condition. One explanation is that the expected benefits of the Sharing condition, namely sustained engagement through information sharing and strengthened grandparent--grandchild relationships, likely require a longer period to accumulate before exerting measurable effects on mental health. Furthermore, mental health is influenced by multiple contextual factors and exhibits substantial daily variability, making short-term changes difficult to attribute to condition-specific mechanisms \citep{Gloster2017-bl}. Future work should therefore examine how sharing personal information influences engagement, relationship dynamics, and ultimately mental health over time. A longitudinal study spanning several months to a year would enable a more precise assessment of these cumulative effects and provide deeper insight into how such communication contributes to mental health.

\subsection{Limitations and Future Work}
As described in Section~\ref{sec:dialogue_design}, the dialogue provided by the proposed agent was intentionally designed to be short and lightweight in order to reduce cognitive and psychological burden for users, and to facilitate sustained engagement over time. However, prior research has shown that conversations involving deeper self-disclosure can play an important role in fostering human--agent relationships \citep{Bickmore2005-vb,Mitsuno2026-rx}, and that sharing more private information may further enhance interpersonal relationships between users \citep{Altman1973-hv}. Given this, it is possible that incorporating more personal or emotionally rich topics could provide additional relational or emotional benefits. At the same time, introducing such content may increase users' cognitive and psychological burden, suggesting the need to carefully balance these aspects in dialogue design. Therefore, future work should explore how to incorporate not only everyday events but also more private aspects of users' lives, while maintaining usability, and examine the effects of such designs.

However, enabling conversations about private matters introduces additional challenges. Because the proposed agent shares information from one user with the other, privacy protection becomes increasingly critical when handling sensitive content. In the present study, participants were informed in advance that some of their messages might be shared with their partner, and consent was obtained accordingly. Nevertheless, when dealing with more personal topics, it will be essential to provide mechanisms that allow users to exercise granular control over what information is shared, ensure transparency in the sharing process, and prevent accidental or inappropriate disclosure due to contextual misunderstanding. Hence, developing systems that robustly support these requirements for secure, user-controlled, and transparent information sharing will be an important direction for future work.

Furthermore, the experimental period in this study lasted only ten days, which may have been insufficient to fully observe changes such as strengthened relationships or improvements in mental health. These forms of psychological change typically accumulate gradually and may not manifest within a short time frame \citep{Gloster2017-bl}. This limitation may partly explain why certain subjective measures, such as IOS and Depression, did not show significant changes. Longer-term longitudinal studies, spanning several months to a year, will be valuable for elucidating how information sharing through dialogue agents influences interpersonal relationships and mental health over time.

\section{Conclusion} \label{sec:conclusion}
This study proposed a dialogue agent designed not only to support human--agent interaction but also to facilitate human--human relationships between users by sharing personal information. The agent operates on a chatbot platform and engages in daily conversations with both older adults and their grandchildren, sharing everyday information obtained from one user with the other.

We conducted a ten-day field experiment with 104 grandparent--grandchild pairs to evaluate the effectiveness of the proposed agent. The results showed that, from a behavioral perspective, older adults became more willing to interact with the agent when it shared information about their grandchild. In addition, based on open-ended responses, the perceived connection between grandparents and grandchildren showed positive changes in the Sharing condition. Moreover, daily interactions with the agent were associated with reduced anxiety in both generations.

Taken together, these findings indicate that a dialogue agent that shares personal information can be an effective approach for supporting older adults by simultaneously offering conversational opportunities and promoting family connectedness. Future work should investigate how long-term use shapes relationship development and mental health trajectories, and further explore design strategies that enable safe, sustained, and meaningful information sharing.

\section*{Declarations}
\bmhead{Funding}
This work was supported by the JST Moonshot R\&D Program (Grant Number JPMJMS2011) for system development and JSPS KAKENHI (Grant Number 24H00165) for data analysis.

\bmhead{Data Availability}
Data will be made available on reasonable request.

\bmhead{Conflict of interest}
The authors declare that they have no conflict of interest.





\bibliography{bibtex}

\end{document}